\documentclass[prl,aps,a4paper,twocolumn,superscriptaddress]{revtex4-1}

\usepackage[utf8x]{inputenc}
\usepackage{amssymb}
\usepackage{amsmath}
\usepackage{graphicx}
\usepackage{bbm}
\usepackage{psfrag}
\usepackage{latexsym}
\usepackage{hyperref}
\usepackage{color}

\usepackage{amsfonts}
\usepackage{amsmath}
\allowdisplaybreaks[4]        
\usepackage{amssymb}
\usepackage{euscript}     
\usepackage{color}  
\usepackage{xcolor}       
\usepackage{tensor}        
\usepackage{amsthm}
\usepackage{graphicx}
\usepackage{subfigure}
\usepackage{bbm}
\usepackage[header,title,page,titletoc]{appendix}

\makeatletter 
\def\@eqnnum{{\normalsize \normalcolor (\theequation)}} 
\makeatother

\newcommand{\be}{\begin{equation}}
\newcommand{\ee}{\end{equation}}
\newcommand{\bea}{\begin{eqnarray}}
\newcommand{\eea}{\end{eqnarray}}

\newcommand{\zbar}{{\bar z}}

\newcommand{\OO}{{\cal O}}

\newcommand{\ra}{{\; \rightarrow \; }}
\newcommand{\GG}{{\cal G}}
\newcommand{\FF}{{\cal F}}
\newcommand{\DL}{{\Delta_L}}
\newcommand{\OL}{{\cal O}_L}
\newcommand{\OH}{{\cal O}_H}
\newcommand{\dxm}{{\Delta x^-}}
\newcommand{\dxp}{{\Delta x^+}}
\newcommand{\tdxm}{{\Delta {\tilde x}^-}}
\newcommand{\bara}{{\bar \alpha}}

\begin{document}

\author{Andrei Parnachev}\email{parnachev [at] maths.tcd.ie}

\affiliation{\it School of Mathematics and Hamilton Mathematics Institute, Trinity College Dublin, Dublin 2, Ireland}

\title{Near Lightcone  Thermal  Conformal Correlators and Holography}

\begin{abstract}
Heavy-heavy-light-light (HHLL) correlators of pairwise identical scalars in CFTs with a large central charge in any number of dimensions 
admit a  double scaling limit where  the ratio of the heavy conformal dimension to the central charge becomes  large as 
the separation between the light operators becomes null.
In this limit the stress tensor sector of a generic HHLL  correlator receives contributions from the multi stress tensor operators 
with any number of stress tensors, as long as their twist is not increased by  index contractions.
We show how one can compute this leading twist stress tensor sector
when the conformal dimension of the light operators is large
and the stress tensor sector approximates the  thermal CFT correlator.
In this regime the value of the correlator is related to  the length of the  spacelike geodesic  
 which approaches the boundary of the dual asymptotically AdS spacetime at the points of light operator insertions.
We provide a detailed description of the infinite volume limit.  In two spacetime dimensions the HHLL Virasoro vacuum block is reproduced, while in 
four spacetime dimensions  the result is written in terms of  elliptic integrals.
\end{abstract}

\maketitle

\noindent {\bf  1.~Introduction:}
In  $d$-dimensional CFTs with a large central charge $C_T$ there are multi-stress tensor operators composed out of  stress-tensors and derivatives.
The  contribution of such composite operators  to a four-point function forms the stress-tensor sector of this correlator.
The case where the correlator includes two pairwise identical operators $\OO_L$ with  conformal dimension $\DL$ of order unity 
and two identical  heavy operators $\OO_H$ with  conformal dimension $\Delta_H\sim C_T$ (with the ratio $\mu \sim \Delta_H/C_T$ fixed) has  recently been studied in various ways 
 \cite{Kulaxizi:2018dxo,Fitzpatrick:2019zqz,Karlsson:2019qfi,Li:2019tpf,Kulaxizi:2019tkd,Fitzpatrick:2019efk,Karlsson:2019dbd,Li:2019zba,Karlsson:2019txu,Karlsson:2020ghx,Li:2020dqm}.

Such heavy-heavy-light-light (HHLL) correlators can be used to determine the OPE coefficients of the two light operators with the multi-stress tensors;  with this data 
one can compute the stress tensor sector of any correlator.
In CFTs with holographic duals  \cite{Maldacena:1997re,Witten:1998qj,Gubser:1998bc}
we expect generic  heavy states created by $\OO_H$ to thermalize;
thermal holographic CFTs are described by asymptotically AdS black holes \cite{Witten:1998zw}.
(For recent work on thermalization in CFTs see e.g. \cite{Lashkari:2016vgj,Lashkari:2017hwq,Dymarsky:2019etq}.)
Hence, the HHLL correlators provide a window into thermal correlators in CFTs.

It will be convenient to place the CFT on a lorentzian cylinder whose base is  a ($d{-}1$)-dimensional sphere of radius $R$.
Consider a kinematical setup where  the heavy operators create a heavy state in the infinite past (and annihilate it in the future)
while the light operators approach each other's lightcone from the spatial direction.
In the language of conformal cross-ratios $z, \zbar$, this corresponds to taking ($1{-}\zbar$) to zero, while keeping $z$ fixed.
In this case the main contribution to the $\OO(\mu^k)$ term in the stress tensor sector  comes from the
multi stress operators with $k$ stress tensors with minimal twist $\tau = k (d-2)$.
An interesting double-scaling limit involves taking $\mu$ large, keeping $\dxm \equiv i \mu^\frac{2}{d-2} (1-\zbar)$ fixed \footnote{
As explained in Section 2, $\dxm$ is the  separation between the insertions of the light operators in the rescaled lightlike direction.}. 
In this limit all minimal twist multi stress tensor operators contribute.

In \cite{Fitzpatrick:2019zqz} it was argued, in a holographic setting, that the OPE coefficients of such operators with two light scalars are universal,
i.e. do not depend on the gravitational lagrangian in the bulk.
A procedure for computing these OPE coefficients using lightcone conformal bootstrap \cite{Fitzpatrick:2012yx,Komargodski:2012ek}
was proposed in \cite{Karlsson:2019dbd}.
At each twist there are infinitely many multi stress tensors labeled by  spin, but 
one can perform the summation  and obtain a simple expression \cite{Kulaxizi:2019tkd}. 
As argued in  \cite{Karlsson:2019dbd}, the leading twist part of the stress tensor sector  exponentiates:
\be\label{expo}
       \langle \OH(x_4)\OL(1)\OL(z,\zbar)\OH(0)\rangle \sim e^ {\DL \FF},
\ee      
where $ \FF=\FF(\dxm,z,\DL)$ is a universal  function  which can be computed order by order in $ (\dxm)^\frac{d-2}{2}$ \cite{Karlsson:2019dbd} 
and has a finite limit as $\DL\ra\infty$, which we denote by $\FF_\infty$.
Hence, the situation is very similar to that in two spacetime dimensions, where the HHLL Virasoro vacuum block 
also has the form (\ref{expo}).
The HHLL Virasoro block was computed exactly 
in a variety of ways (see e.g. \cite{Fitzpatrick:2014vua,Fitzpatrick:2015zha,Hijano:2015rla,Hijano:2015qja,Fitzpatrick:2015foa,Cotler:2018zff,Collier:2018exn}).

The situation in $d>2$ is much more intricate and the function $\FF$ is not known.
In this paper we use holography to compute $\FF_\infty$.
Note that in the $\DL\gg 1$ limit the multi-trace operators which include one or more insertion of $\OO_L$  decouple and the 
stress tensor sector approximates the full thermal correlator, at least for holographic theories.
The value of $\FF_\infty$ equals minus the (regularized) length of the spacelike geodesic which propagates in 
the $d+1$-dimensional AdS-Schwarzschild background and 
approaches  the positions  of the light operator insertions at the boundary.
This  has  recently been used to partially compute the multi stress tensor OPE coefficients \cite{Fitzpatrick:2019zqz,Kulaxizi:2019tkd} (see also \cite{Maxfield:2017rkn}).
We pay particular attention to the large volume  limit, where $R\ra \infty$ and $(1-z) \ra 0$.
In this case the function $\FF_\infty$ simplifies and we provide direct comparison (and exact matching) for the first few terms in the expansion. 

The rest of the paper is organized as follows.
In the next Section we provide a brief review and set up notation along the way.
In Section  3 we compute the effective metric where the geodesic, whose length computes  the 
near lightcone correlator, propagates.
We also write down   expressions which allow the computation of $\FF_\infty$.
In Section 4 we consider the large volume limit and reproduce the Virasoro vacuum block in $d=2$.
We also compute $\FF_\infty$ in the four-dimensional case  and match the first few terms in the expansion in $\dxm$ 
with the known results.
We discuss the results in Section 5.
Appendix A contains some details related to the $d=2$ case, while Appendix B describes the $d=4$ case.

\bigskip
\noindent {\bf 2.~Review:}
We will consider the stress-tensor sector of the correlator, where the heavy operators $\OO_H$ are inserted at $t=\pm\infty$,
and the relative coordinates of the light operators differ by $\Delta t$ in the time direction and by $\Delta \varphi$ on the ($d{-}1$)-sphere.
The contribution of the multi stress-tensors (denoted by the subscript "MST") can be written as
\be
\label{Corrdef}
	\GG(z,\zbar) = \lim_{x_4\to\infty} x_4^{2\Delta_H}\langle \OH(x_4)\OL(1)\OL(z,\zbar)\OH(0)\rangle_{\rm MST} .
\ee
The conformal dimension of the heavy operators, $\OO_H$, is proportional to the central charge with 
\be
\label{defmu}
\mu \equiv \frac{4 \Gamma(d+2)}{(d-1)^2 \Gamma(\frac{d}{2})^2} \frac{\Delta_H}{C_T}
\ee
fixed.
The cross-ratios $z$ and $\zbar$ are related to the relative spacetime positions of the light operators as (see e.g. \cite{Kulaxizi:2018dxo})
\be
\label{zzbar}
  z= \exp( i \Delta x^+ ), \;  \zbar= \exp( i \Delta \tilde x^- ),
\ee  
where
\be\label{dxpmtphi}  
   \Delta x^+ \equiv (\Delta t + \Delta \varphi), \; \Delta \tilde x^- \equiv (\Delta t - \Delta\varphi).
 \ee
 In the equations above, and in most of what follows we set $R=1$.
To retain the  contributions of  all leading twist multi stress tensors we take the double-scaling limit $\mu \ra \infty, \; (1- \bar z) \ra 0$
with 
\be
\label{defdxm}
       \dxm \equiv \mu^\frac{2}{d-2} \, \Delta \tilde x^-   \approx  i  \mu^\frac{2}{d-2} (1- \bar z)
\ee
fixed.
The double scaling limit of the stress-tensor sector of the HHLL correlator, which we denote by $\tilde \GG$,  exponentiates,
\be
\label{corr}
   \tilde \GG = \exp( \DL \FF ), \qquad \FF = \sum_{k=1}^\infty   \mu^k (1- \zbar)^\frac{(d-2)k}{2} \FF^{(k)} .
 \ee
In (\ref{corr}) $\FF^{(k)} = \FF^{(k)} (z,\DL)$ receives contributions from the 
 minimal twist multi-stress tensor operators with $k$ stress tensors.
 In $d=2$ the function $\FF$ is $\DL$-independent, 
 \be
 \label{vir}
    \FF|_{d=2} = - \log \sinh \left( \frac{\sqrt{\mu-1} }{2}  \dxp \right).
 \ee
 In higher dimensions it is possible to compute $\FF^{(k)} $ using conformal bootstrap, order by order in $k$.
 The $k=1$ terms is just the stress tensor contribution, 
$\FF^{(1)} \sim f_\frac{d+2}{2}(z)$  where $f_a(z) \equiv (1-z)^a\; _2F_1(a,a,2 a, 1-z)$.
The $k=2$ term is a result of the summation of all twist-$2(d-2)$ double stress tensors
with varying spin (one can find the explicit expression for $d=4$ in \cite{Kulaxizi:2019tkd}).
It will be useful to consider the $\DL \ra\infty$ limit of the $k=2$ term,
\be
\label{f1lim}
  \FF^{(2)}_\infty|_{d=4} = {-5 f_3(z)^2+{15\over 7} f_2(z)f_4(z)+{40\over 7}  f_1(z)f_5(z)\over 28800}.
\ee
The next, $k=3$ term in the expansion, was computed using bootstrap in \cite{Karlsson:2019dbd}.
To take the large volume limit one needs to set the hypergeometric functions in $f_a(z)$ to unity.
(This corresponds to keeping only multi stress tensors without derivatives).
The result is
\be
\label{ff}
\begin{split}
  \FF_\infty|_{d=4} & \simeq {-}\log( \dxp \dxm) {+} \frac{ \dxm (\dxp)^3}{120}  \\
     & \;\; +   \frac{ (\dxm)^2 (\dxp)^6}{10080} {+}   \frac{ 1583 (\dxm)^3 (\dxp)^9}{648648000}+\ldots,
  \end{split}
 \ee
 where we used (\ref{zzbar}) to write the result in terms of $\dxp, \dxm$.
 The first term in the right hand side  of (\ref{ff}) corresponds to the vacuum two-point function.
 
\bigskip
\noindent {\bf 3.~Effective metric and spacelike geodesics:}
 We would like to analyze spacelike geodesics in   ($d{+}1$)-dimensional
 AdS-Schwarzschild spacetime:
 \be
 \label{AdSS}
 ds^2 =-f_{BH}  dt^2 + f_{BH}^{-1} dr^2 +r^2 (d \varphi^2 +\sin^2 \varphi d\Omega_{d-2})^2,
 \ee
 where 
 \be
 \label{fbh}
   f_{BH} = 1+ r^2 -\frac{\mu}{r^{d-2}} .
 \ee
 As discussed above, we are interested in the limit $\mu\ra\infty$.
 We will consider  spacelike geodesics which
approach the AdS boundary at  points separated in the lightlike direction  $\tilde x^- = t-\varphi$ 
by  $\Delta \tilde x^ -\ra 0$ with $\dxm = \mu^\frac{2}{d-2} \tdxm$ fixed \footnote{ In the following we 
mostly consider $d>2$, but the $d=2$ limit can also be recovered.}.
The geodesics can be taken to live in the $t, \varphi, r$ part of the spacetime.
It is natural to  write the metric  (\ref{AdSS}) in  coordinates   $x^+$, $ x^- =  \tilde x^- \mu^\frac{2}{d-2}$ and $y =r \mu^{-\frac{1}{d-2}}$.
In the $\mu\ra\infty$ limit the metric of the 
three-dimensional spacetime where the geodesics propagate becomes
\be
\label{metlimit}
   ds^2 = -\frac{1}{4} \left(1- \frac{1}{y^{d-2}} \right) (dx^+)^2 -y^2 dx^+ \, d x^- +\frac{dy^2}{y^2}.
\ee
 The two Killing vectors give rise to two conserved quantities,
 \be
 \label{conserved}
 K_+{=}  -\frac{1}{4} \left(1- \frac{1}{y^{d-2}} \right) \dot x^+ -\frac{y^2}{2} \dot x^-, \;
 K_-{=}-K{=}-\frac{y^2}{2} \dot x^+ .
 \ee
 The geodesic equation becomes
 \be\label{geo}
   \dot y^2 + 4 K K_+ +  (y^{-2}- y^{-d} ) K^2 -y^2=0 .
 \ee
Eq. (\ref{geo}) describes the one-dimensional motion of a particle in an effective potential
 which can be inferred from (\ref{geo}).
An important quantity is the largest (real) solution of the equation 
 \be\label{turning}
    4 K K_+ +  (y_0^{-2}- y_0^{-d} ) K^2 -y_0^2=0.
 \ee
 It specifies the turning point of the particle. 
 Now one can compute the length of the geodesic $\ell$, as well as $\dxp$ and $\dxm$, in terms of $K$, $K_+$:
 \be\label{dxp_int}
 \dxp = 4 K  \int_{y_0}^\infty \frac{dy}{ y^2 \left(   (y^{-d}-y^{-2}) K^2 -4 K K_+ +y^2  \right)^\frac{1}{2}  } ,
 \ee
 \be\label{dxm_int}
 \dxm = 2    \int_{y_0}^\infty dy \frac{ -2 K_+    +   (y^{-d}-y^{-2}) K }{ y^2 \left(   (y^{-d}-y^{-2}) K^2 -4 K K_+ +y^2  \right)^\frac{1}{2}  } ,
 \ee
 \be\label{ell_int}
 \ell = 2  \int_{y_0}^\infty \frac{dy}{ \left(   (y^{-d}-y^{-2}) K^2 -4 K K_+ +y^2  \right)^\frac{1}{2}  } .
 \ee
 Rewriting  $-\ell$ in terms of $\dxp$ and $\dxm$ yields the value of $\FF_\infty$.
 
 Let us consider the large $K$ behavior.
 The substitution $y = \sqrt{K} \tilde y$ can be  used to argue 
that in this limit   the $y^{-d} K^2$  term under the square root in the denominators of (\ref{dxp_int})- (\ref{ell_int}) can be dropped.
The subsequent integration yields $\dxp = 2 \cot ^{-1} (2 K_+)$ and  $\dxm =  - K^{-1} (K_+ + \frac{1}{2} ) \dxp$.
 Now we see that the large $K$ limit corresponds to the $\dxm \ra 0$ limit. 
 We can now compute  the length and recover $\FF_\infty$, but the immediate technical difficulty is that
 the length is divergent.
 This means we need to regularize it -- this corresponds to introducing  a UV cutoff in the dual CFT.
 It will be easier to do this  in the large volume limit.

\bigskip
\noindent {\bf 4.~ Large volume limit:}
In the following we simplify the setup further and take the large volume limit (CFT on $\mathbb{R}^{d-1,1}$, as opposed to $\mathbb{S}^{d-1}\times \mathbb{R}$ ).
This can be achieved in two equivalent ways: either taking the limit $\dxm \sim R^{d+2\over d-2} \ra\infty$, $\dxp \sim R^{-1}\ra 0$
(which corresponds to taking $K_+\sim R\ra \infty$, $K\sim R^{-{d+2\over d-2}} \ra 0$) or
dropping the unity in the $g_{tt}$ metric component in (\ref{metlimit}), which describes the asymptotically AdS black hole with a planar horizon.

Either way, in this limit the integrals become
 \be\label{dxp_planar}
 \dxp \simeq 4 K  \int_{y_0}^\infty \frac{dy}{ y^2 \left(   y^{-d} K^2 -4 K K_+ +y^2  \right)^\frac{1}{2}  } ,
 \ee
 \be\label{dxm_planar}
 \dxm \simeq 2    \int_{y_0}^\infty dy \frac{ -2 K_+    +   y^{-d} K }{ y^2 \left(   y^{-d}K^2 -4 K K_+ +y^2  \right)^\frac{1}{2}  } ,
 \ee
 \be\label{ell_planar}
 \ell_\Lambda \simeq 2  \int_{y_0}^{\Lambda} \frac{dy}{ \left(   y^{-d} K^2 -4 K K_+ +y^2  \right)^\frac{1}{2}  } .
 \ee
 where the subscript $\Lambda$ has been introduced to signify the cutoff dependence.
 In the integrals above the lower limit of integration corresponds to the largest root 
 of the expression inside the square root and ``$\simeq$" means equality up to terms subleading in $1/R$.
 It is useful to rewrite the expressions above as
\be\label{dxp_planara}
\dxp {\simeq}  \frac{1}{K_+}  I_+(\alpha), \; 
I_+(x) \equiv 4 \int_{u_0}^\infty \frac{ du}{u^{\frac{4-d}{2}} \left( u^{d+2}{-}4 u^d +x \right)^\frac{1}{2} } ,
 \ee
\be\label{dxm_planara}
\dxm {\simeq}  -\frac{1}{K}  I_-(\alpha), \; 
I_-(x) \equiv 4 \int_{u_0}^\infty \frac{\left( 1- \frac{x}{2 u^d}  \right) du}{u^{\frac{4-d}{2}} \left( u^{d+2}{-}4 u^d +x \right)^\frac{1}{2} } , 
 \ee
 \be\label{ell_planara}
 \begin{split}
\ell_f  & \simeq  -\log(K_+ K) +  I_\ell(\alpha), \\
           & I_\ell(x) \equiv 2 \int_{u_0}^{\Lambda_u}   \frac{ u^\frac{d}{2} du }{ (u^{d+2}{-}4 u^d +x )^\frac{1}{2}  } -2 \log \Lambda_u ,
\end{split}
 \ee
 where $\alpha = K^{-\frac{d-2}{2}} K_+^{-\frac{d+2}{2}}$ and the subscript ``$f$" in (\ref{ell_planara})
 stands for the regularized value.
 The appearance of the $\log(K_+ K) $ term in (\ref{ell_planara}) comes from the regularization 
 and the change of variables  $u =(K K_+)^{-1/2} y$ -- this implies that the original cutoff $\Lambda$
 is related to  $\Lambda_u$ via $\Lambda= (K K_+)^{1/2} \Lambda_u$.
 In (\ref{ell_planara}) the limit $\Lambda_u\ra\infty$ is implied; in this limit the expression for 
 $ I_\ell(x) $ is cutoff-independent.
 
 Here and in what follows we set the regularization-dependent  constant term to zero (this corresponds to 
the canonical normalization of the conformal $\langle \OO_L \OO_L\rangle$  two point function).
Combining everything,
 \be
\label{ellcutoff_f}
 \ell_f \simeq   \log (\dxp\dxm )  +\delta I ,
\ee
where 
\be
\label{deltai}
\delta I = -\log[ I_+(\alpha)  I_-(\alpha) ] +I_\ell(\alpha)
\ee
determines the correlator with the vacuum part subtracted 
and $\alpha$ is a solution of
\be
\label{alphasol}
   (-\dxm)^{\frac{d-2}{2}} (\dxp)^{\frac{d+2}{2}} = \alpha \, I_-^{\frac{d-2}{2}}(\alpha) \, I_+^{\frac{d+2}{2}}(\alpha) .
\ee

The expressions above provide a complete solution in the large volume limit,
although one still needs to compute (and invert) a few functions which are determined
by simple one-dimensional integrals.

 To extract the $d=2$ result, we need to take the $d\ra2$ limit of (\ref{alphasol}) keeping $\Delta \tilde x^-$ fixed.
 This yields $\sqrt{\mu} \dxp = \alpha I_+^{(d=2)}(\alpha)$.
 One can further show  $I_-^{(d=2)}(\alpha)=1$ and (see Appendix for details)
 \be
 \label{dtwo}
     -\ell_f|_{d=2} \simeq -\log \sinh \frac{\sqrt{\mu} \dxp}{2}  ,
 \ee
 which agrees with (\ref{vir}) in the large volume limit, $\mu\gg 1$.

 \begin{figure}
	\includegraphics[width=3.475in]{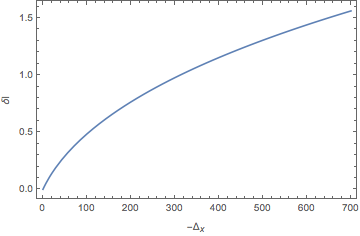}
	\caption{\label{deltaiplot} The plot of $\delta I$, which exponentiates
	to produce the HHLL correlator with the vacuum part subtracted  [see eq.  (\ref{deltai})], in $d=4$ as a function of $|\Delta_x|$.}
\end{figure}

Likewise,  we can substitute $d=4$ in the  general expressions above and obtain the solution
in terms of elliptic integrals (see Appendix for details).
The function $\delta I$ [see eq. (\ref{deltai})] is plotted in Fig. 1.
 One can also expand to any desired order in $\Delta_x \equiv \dxm (\dxp)^3$:
 \be
\label{ellcutoff_f}
\begin{split}
    -\ell_f|_{d=4} & \simeq  {-}\log (\dxm \dxp) {+ }\frac{ \Delta_x}{120}  +   \frac{ \Delta_x^2}{10080} +   \frac{ 1583\Delta_x^3}{648648000}\\
    & \qquad +   \frac{3975313 \Delta_x^4}{49401031680000} +\ldots .
    \end{split}
\ee
Terms up to $\OO(\Delta_x^3)$ agree with the previously known result (\ref{ff}).
Note that this is the lightcone limit of the small temperature expansion of the correlator
(in the large volume limit $\Delta_x =c\,  T^4 \dxm (\dxp)^3$ where $T$ is the temperature, $c$ is a
theory-dependent numerical coefficient, and factors of $R$ have explicitly canceled out).

\bigskip
\noindent {\bf 5.~Discussion:}
We have computed the near lightcone behavior of  HHLL correlators
in CFTs with a large central charge in the large $\DL$ limit.
The result provides a close analog of the Virasoro vacuum block in large-$C_T$ two-dimensional CFTs.
It would  be nice to see if the $d=4$ result could be 
simplified further (while we do have an expression in terms of the Appell functions/elliptic integrals, it is still a bit involved).
Perhaps this would allow understanding of the algebraic structure behind the near-lightcone correlators
(see \cite{Huang:2019fog,Huang:2020ycs} for recent work in this direction).

We have explicitly matched the expansion of the length of the near-null geodesic to the
near lightcone behavior of the  correlator in the large $R$ and large $\DL$ limit.
Note that one can read off the OPE coefficients from the exponential of $\FF_\infty$. 
For example, the second term in  the right-hand side of (\ref{ff})  determines 
 the OPE coefficient of the two scalars $\OO_L$ with the stress-tensor  \emph{and} the  leading [$\OO(\DL^k)$] 
 behavior of the OPE coefficients of   $\OO_L$s with the leading twist $k$-stress tensor operator $T_{\mu\nu}^k$.

One may ask whether the full correlator is well approximated by its stress-tensor sector.
Generally we expect the multi trace operators of
the form $[\OO_L^p T_{\mu\nu}^q]_{n,l}$,  to contribute.
In the $d=2$ case, such contributions  vanish in the large $R$ limit  and the HHLL Virasoro vacuum block  
agrees with the full thermal correlator in the high temperature limit
(in this limit the correlator can be computed by using a conformal transformation from $\mathbb{R}^2$ to $\mathbb{R}\times \mathbb{S}^1$).
The situation in higher dimensions is more intricate (see e.g.  \cite{Fitzpatrick:2019zqz} for a recent discussion)
 \footnote{ Note
that multi stress tensor operators 
with any derivatives in them do not contribute in the large $R$ limit.}.
Nevertheless, in the large $\DL$ limit, multi-trace operators involving $\OO_L$ become heavy 
and decouple even for finite $R$.
Hence, in the large $\DL$ limit the stress tensor sector is a good approximation to  the thermal correlator.

The results of Section 4 make it evident  that as  $\alpha$ approaches a certain critical value, the
spacelike geodesic gets closer to the horizon (this corresponds to the $\dxp \ra\infty$ limit;
similar behavior of geodesics was recently studied in \cite{Bianchi:2020des}).
It would be interesting to investigate this limit
from the CFT point of view.
It would also be interesting to relate our results to the behavior of quasinormal modes in the UV
region \cite{Brigante:2007nu,Brigante:2008gz}
 which is relevant for the conformal collider bounds \cite{Hofman:2008ar}.

Other natural  directions include 
explicit computations for finite $R$ (this would involve  keeping the hypergeometric functions in (\ref{ff}) ) 
and  generalization for finite values of $\DL$.
Also note that the universality of the multi stress tensor OPE coefficients may have a wider region of applicability than just the set
of holographic theories.
In this case the holographic calculation of this paper will have a wider regime of applicability as well.


\bigskip
\noindent{\bf Acknowledgements:}
I benefited from discussions with R. Karlsson, M. Kulaxizi, G.S. Ng, K. Sen and P. Tadi\'c.
This work was supported in part by an Irish Research Council  Consolidator Laureate Award.

\bibliography{lcthv2}

\begin{thebibliography}{37}%
\makeatletter
\providecommand \@ifxundefined [1]{%
 \@ifx{#1\undefined}
}%
\providecommand \@ifnum [1]{%
 \ifnum #1\expandafter \@firstoftwo
 \else \expandafter \@secondoftwo
 \fi
}%
\providecommand \@ifx [1]{%
 \ifx #1\expandafter \@firstoftwo
 \else \expandafter \@secondoftwo
 \fi
}%
\providecommand \natexlab [1]{#1}%
\providecommand \enquote  [1]{``#1''}%
\providecommand \bibnamefont  [1]{#1}%
\providecommand \bibfnamefont [1]{#1}%
\providecommand \citenamefont [1]{#1}%
\providecommand \href@noop [0]{\@secondoftwo}%
\providecommand \href [0]{\begingroup \@sanitize@url \@href}%
\providecommand \@href[1]{\@@startlink{#1}\@@href}%
\providecommand \@@href[1]{\endgroup#1\@@endlink}%
\providecommand \@sanitize@url [0]{\catcode `\\12\catcode `\$12\catcode
  `\&12\catcode `\#12\catcode `\^12\catcode `\_12\catcode `\%12\relax}%
\providecommand \@@startlink[1]{}%
\providecommand \@@endlink[0]{}%
\providecommand \url  [0]{\begingroup\@sanitize@url \@url }%
\providecommand \@url [1]{\endgroup\@href {#1}{\urlprefix }}%
\providecommand \urlprefix  [0]{URL }%
\providecommand \Eprint [0]{\href }%
\providecommand \doibase [0]{http://dx.doi.org/}%
\providecommand \selectlanguage [0]{\@gobble}%
\providecommand \bibinfo  [0]{\@secondoftwo}%
\providecommand \bibfield  [0]{\@secondoftwo}%
\providecommand \translation [1]{[#1]}%
\providecommand \BibitemOpen [0]{}%
\providecommand \bibitemStop [0]{}%
\providecommand \bibitemNoStop [0]{.\EOS\space}%
\providecommand \EOS [0]{\spacefactor3000\relax}%
\providecommand \BibitemShut  [1]{\csname bibitem#1\endcsname}%
\let\auto@bib@innerbib\@empty
\bibitem [{\citenamefont {Kulaxizi}\ \emph
  {et~al.}(2019{\natexlab{a}})\citenamefont {Kulaxizi}, \citenamefont {Ng},\
  and\ \citenamefont {Parnachev}}]{Kulaxizi:2018dxo}%
  \BibitemOpen
  \bibfield  {author} {\bibinfo {author} {\bibfnamefont {M.}~\bibnamefont
  {Kulaxizi}}, \bibinfo {author} {\bibfnamefont {G.~S.}\ \bibnamefont {Ng}}, \
  and\ \bibinfo {author} {\bibfnamefont {A.}~\bibnamefont {Parnachev}},\ }\href
  {\doibase 10.21468/SciPostPhys.6.6.065} {\bibfield  {journal} {\bibinfo
  {journal} {SciPost Phys.}\ }\textbf {\bibinfo {volume} {6}},\ \bibinfo
  {pages} {065} (\bibinfo {year} {2019}{\natexlab{a}})},\ \Eprint
  {http://arxiv.org/abs/1812.03120} {arXiv:1812.03120 [hep-th]} \BibitemShut
  {NoStop}%
\bibitem [{\citenamefont {Fitzpatrick}\ and\ \citenamefont
  {Huang}(2019)}]{Fitzpatrick:2019zqz}%
  \BibitemOpen
  \bibfield  {author} {\bibinfo {author} {\bibfnamefont {A.~L.}\ \bibnamefont
  {Fitzpatrick}}\ and\ \bibinfo {author} {\bibfnamefont {K.-W.}\ \bibnamefont
  {Huang}},\ }\href {\doibase 10.1007/JHEP08(2019)138} {\bibfield  {journal}
  {\bibinfo  {journal} {JHEP}\ }\textbf {\bibinfo {volume} {08}},\ \bibinfo
  {pages} {138} (\bibinfo {year} {2019})},\ \Eprint
  {http://arxiv.org/abs/1903.05306} {arXiv:1903.05306 [hep-th]} \BibitemShut
  {NoStop}%
\bibitem [{\citenamefont {Karlsson}\ \emph {et~al.}(2019)\citenamefont
  {Karlsson}, \citenamefont {Kulaxizi}, \citenamefont {Parnachev},\ and\
  \citenamefont {Tadi\'c}}]{Karlsson:2019qfi}%
  \BibitemOpen
  \bibfield  {author} {\bibinfo {author} {\bibfnamefont {R.}~\bibnamefont
  {Karlsson}}, \bibinfo {author} {\bibfnamefont {M.}~\bibnamefont {Kulaxizi}},
  \bibinfo {author} {\bibfnamefont {A.}~\bibnamefont {Parnachev}}, \ and\
  \bibinfo {author} {\bibfnamefont {P.}~\bibnamefont {Tadi\'c}},\ }\href
  {\doibase 10.1007/JHEP10(2019)046} {\bibfield  {journal} {\bibinfo  {journal}
  {JHEP}\ }\textbf {\bibinfo {volume} {10}},\ \bibinfo {pages} {046} (\bibinfo
  {year} {2019})},\ \Eprint {http://arxiv.org/abs/1904.00060} {arXiv:1904.00060
  [hep-th]} \BibitemShut {NoStop}%
\bibitem [{\citenamefont {Li}\ \emph {et~al.}(2019)\citenamefont {Li},
  \citenamefont {Mai},\ and\ \citenamefont {Lu}}]{Li:2019tpf}%
  \BibitemOpen
  \bibfield  {author} {\bibinfo {author} {\bibfnamefont {Y.-Z.}\ \bibnamefont
  {Li}}, \bibinfo {author} {\bibfnamefont {Z.-F.}\ \bibnamefont {Mai}}, \ and\
  \bibinfo {author} {\bibfnamefont {H.}~\bibnamefont {Lu}},\ }\href {\doibase
  10.1007/JHEP09(2019)001} {\bibfield  {journal} {\bibinfo  {journal} {JHEP}\
  }\textbf {\bibinfo {volume} {09}},\ \bibinfo {pages} {001} (\bibinfo {year}
  {2019})},\ \Eprint {http://arxiv.org/abs/1905.09302} {arXiv:1905.09302
  [hep-th]} \BibitemShut {NoStop}%
\bibitem [{\citenamefont {Kulaxizi}\ \emph
  {et~al.}(2019{\natexlab{b}})\citenamefont {Kulaxizi}, \citenamefont {Ng},\
  and\ \citenamefont {Parnachev}}]{Kulaxizi:2019tkd}%
  \BibitemOpen
  \bibfield  {author} {\bibinfo {author} {\bibfnamefont {M.}~\bibnamefont
  {Kulaxizi}}, \bibinfo {author} {\bibfnamefont {G.~S.}\ \bibnamefont {Ng}}, \
  and\ \bibinfo {author} {\bibfnamefont {A.}~\bibnamefont {Parnachev}},\ }\href
  {\doibase 10.1007/JHEP10(2019)107} {\bibfield  {journal} {\bibinfo  {journal}
  {JHEP}\ }\textbf {\bibinfo {volume} {10}},\ \bibinfo {pages} {107} (\bibinfo
  {year} {2019}{\natexlab{b}})},\ \Eprint {http://arxiv.org/abs/1907.00867}
  {arXiv:1907.00867 [hep-th]} \BibitemShut {NoStop}%
\bibitem [{\citenamefont {Fitzpatrick}\ \emph {et~al.}(2019)\citenamefont
  {Fitzpatrick}, \citenamefont {Huang},\ and\ \citenamefont
  {Li}}]{Fitzpatrick:2019efk}%
  \BibitemOpen
  \bibfield  {author} {\bibinfo {author} {\bibfnamefont {A.~L.}\ \bibnamefont
  {Fitzpatrick}}, \bibinfo {author} {\bibfnamefont {K.-W.}\ \bibnamefont
  {Huang}}, \ and\ \bibinfo {author} {\bibfnamefont {D.}~\bibnamefont {Li}},\
  }\href {\doibase 10.1007/JHEP11(2019)139} {\bibfield  {journal} {\bibinfo
  {journal} {JHEP}\ }\textbf {\bibinfo {volume} {11}},\ \bibinfo {pages} {139}
  (\bibinfo {year} {2019})},\ \Eprint {http://arxiv.org/abs/1907.10810}
  {arXiv:1907.10810 [hep-th]} \BibitemShut {NoStop}%
\bibitem [{\citenamefont {Karlsson}\ \emph
  {et~al.}(2020{\natexlab{a}})\citenamefont {Karlsson}, \citenamefont
  {Kulaxizi}, \citenamefont {Parnachev},\ and\ \citenamefont
  {Tadi\'c}}]{Karlsson:2019dbd}%
  \BibitemOpen
  \bibfield  {author} {\bibinfo {author} {\bibfnamefont {R.}~\bibnamefont
  {Karlsson}}, \bibinfo {author} {\bibfnamefont {M.}~\bibnamefont {Kulaxizi}},
  \bibinfo {author} {\bibfnamefont {A.}~\bibnamefont {Parnachev}}, \ and\
  \bibinfo {author} {\bibfnamefont {P.}~\bibnamefont {Tadi\'c}},\ }\href
  {\doibase 10.1007/JHEP01(2020)076} {\bibfield  {journal} {\bibinfo  {journal}
  {JHEP}\ }\textbf {\bibinfo {volume} {01}},\ \bibinfo {pages} {076} (\bibinfo
  {year} {2020}{\natexlab{a}})},\ \Eprint {http://arxiv.org/abs/1909.05775}
  {arXiv:1909.05775 [hep-th]} \BibitemShut {NoStop}%
\bibitem [{\citenamefont {Li}(2019)}]{Li:2019zba}%
  \BibitemOpen
  \bibfield  {author} {\bibinfo {author} {\bibfnamefont {Y.-Z.}\ \bibnamefont
  {Li}},\ }\href@noop {} {\  (\bibinfo {year} {2019})},\ \Eprint
  {http://arxiv.org/abs/1910.06357} {arXiv:1910.06357 [hep-th]} \BibitemShut
  {NoStop}%
\bibitem [{\citenamefont {Karlsson}(2019)}]{Karlsson:2019txu}%
  \BibitemOpen
  \bibfield  {author} {\bibinfo {author} {\bibfnamefont {R.}~\bibnamefont
  {Karlsson}},\ }\href@noop {} {\  (\bibinfo {year} {2019})},\ \Eprint
  {http://arxiv.org/abs/1912.01577} {arXiv:1912.01577 [hep-th]} \BibitemShut
  {NoStop}%
\bibitem [{\citenamefont {Karlsson}\ \emph
  {et~al.}(2020{\natexlab{b}})\citenamefont {Karlsson}, \citenamefont
  {Kulaxizi}, \citenamefont {Parnachev},\ and\ \citenamefont
  {Tadi\'c}}]{Karlsson:2020ghx}%
  \BibitemOpen
  \bibfield  {author} {\bibinfo {author} {\bibfnamefont {R.}~\bibnamefont
  {Karlsson}}, \bibinfo {author} {\bibfnamefont {M.}~\bibnamefont {Kulaxizi}},
  \bibinfo {author} {\bibfnamefont {A.}~\bibnamefont {Parnachev}}, \ and\
  \bibinfo {author} {\bibfnamefont {P.}~\bibnamefont {Tadi\'c}},\ }\href@noop
  {} {\  (\bibinfo {year} {2020}{\natexlab{b}})},\ \Eprint
  {http://arxiv.org/abs/2002.12254} {arXiv:2002.12254 [hep-th]} \BibitemShut
  {NoStop}%
\bibitem [{\citenamefont {Li}\ and\ \citenamefont {Zhang}(2020)}]{Li:2020dqm}%
  \BibitemOpen
  \bibfield  {author} {\bibinfo {author} {\bibfnamefont {Y.-Z.}\ \bibnamefont
  {Li}}\ and\ \bibinfo {author} {\bibfnamefont {H.-Y.}\ \bibnamefont {Zhang}},\
  }\href@noop {} {\  (\bibinfo {year} {2020})},\ \Eprint
  {http://arxiv.org/abs/2004.04758} {arXiv:2004.04758 [hep-th]} \BibitemShut
  {NoStop}%
\bibitem [{\citenamefont {Maldacena}(1999)}]{Maldacena:1997re}%
  \BibitemOpen
  \bibfield  {author} {\bibinfo {author} {\bibfnamefont {J.~M.}\ \bibnamefont
  {Maldacena}},\ }\href {\doibase 10.1023/A:1026654312961,
  10.4310/ATMP.1998.v2.n2.a1} {\bibfield  {journal} {\bibinfo  {journal} {Int.
  J. Theor. Phys.}\ }\textbf {\bibinfo {volume} {38}},\ \bibinfo {pages} {1113}
  (\bibinfo {year} {1999})},\ \bibinfo {note} {[Adv. Theor. Math.
  Phys.2,231(1998)]},\ \Eprint {http://arxiv.org/abs/hep-th/9711200}
  {arXiv:hep-th/9711200 [hep-th]} \BibitemShut {NoStop}%
\bibitem [{\citenamefont {Witten}(1998{\natexlab{a}})}]{Witten:1998qj}%
  \BibitemOpen
  \bibfield  {author} {\bibinfo {author} {\bibfnamefont {E.}~\bibnamefont
  {Witten}},\ }\href {\doibase 10.4310/ATMP.1998.v2.n2.a2} {\bibfield
  {journal} {\bibinfo  {journal} {Adv. Theor. Math. Phys.}\ }\textbf {\bibinfo
  {volume} {2}},\ \bibinfo {pages} {253} (\bibinfo {year}
  {1998}{\natexlab{a}})},\ \Eprint {http://arxiv.org/abs/hep-th/9802150}
  {arXiv:hep-th/9802150 [hep-th]} \BibitemShut {NoStop}%
\bibitem [{\citenamefont {Gubser}\ \emph {et~al.}(1998)\citenamefont {Gubser},
  \citenamefont {Klebanov},\ and\ \citenamefont {Polyakov}}]{Gubser:1998bc}%
  \BibitemOpen
  \bibfield  {author} {\bibinfo {author} {\bibfnamefont {S.~S.}\ \bibnamefont
  {Gubser}}, \bibinfo {author} {\bibfnamefont {I.~R.}\ \bibnamefont
  {Klebanov}}, \ and\ \bibinfo {author} {\bibfnamefont {A.~M.}\ \bibnamefont
  {Polyakov}},\ }\href {\doibase 10.1016/S0370-2693(98)00377-3} {\bibfield
  {journal} {\bibinfo  {journal} {Phys. Lett.}\ }\textbf {\bibinfo {volume}
  {B428}},\ \bibinfo {pages} {105} (\bibinfo {year} {1998})},\ \Eprint
  {http://arxiv.org/abs/hep-th/9802109} {arXiv:hep-th/9802109 [hep-th]}
  \BibitemShut {NoStop}%
\bibitem [{\citenamefont {Witten}(1998{\natexlab{b}})}]{Witten:1998zw}%
  \BibitemOpen
  \bibfield  {author} {\bibinfo {author} {\bibfnamefont {E.}~\bibnamefont
  {Witten}},\ }\href {\doibase 10.4310/ATMP.1998.v2.n3.a3} {\bibfield
  {journal} {\bibinfo  {journal} {Adv. Theor. Math. Phys.}\ }\textbf {\bibinfo
  {volume} {2}},\ \bibinfo {pages} {505} (\bibinfo {year}
  {1998}{\natexlab{b}})},\ \Eprint {http://arxiv.org/abs/hep-th/9803131}
  {arXiv:hep-th/9803131} \BibitemShut {NoStop}%
\bibitem [{\citenamefont {Lashkari}\ \emph
  {et~al.}(2018{\natexlab{a}})\citenamefont {Lashkari}, \citenamefont
  {Dymarsky},\ and\ \citenamefont {Liu}}]{Lashkari:2016vgj}%
  \BibitemOpen
  \bibfield  {author} {\bibinfo {author} {\bibfnamefont {N.}~\bibnamefont
  {Lashkari}}, \bibinfo {author} {\bibfnamefont {A.}~\bibnamefont {Dymarsky}},
  \ and\ \bibinfo {author} {\bibfnamefont {H.}~\bibnamefont {Liu}},\ }\href
  {\doibase 10.1088/1742-5468/aab020} {\bibfield  {journal} {\bibinfo
  {journal} {J. Stat. Mech.}\ }\textbf {\bibinfo {volume} {1803}},\ \bibinfo
  {pages} {033101} (\bibinfo {year} {2018}{\natexlab{a}})},\ \Eprint
  {http://arxiv.org/abs/1610.00302} {arXiv:1610.00302 [hep-th]} \BibitemShut
  {NoStop}%
\bibitem [{\citenamefont {Lashkari}\ \emph
  {et~al.}(2018{\natexlab{b}})\citenamefont {Lashkari}, \citenamefont
  {Dymarsky},\ and\ \citenamefont {Liu}}]{Lashkari:2017hwq}%
  \BibitemOpen
  \bibfield  {author} {\bibinfo {author} {\bibfnamefont {N.}~\bibnamefont
  {Lashkari}}, \bibinfo {author} {\bibfnamefont {A.}~\bibnamefont {Dymarsky}},
  \ and\ \bibinfo {author} {\bibfnamefont {H.}~\bibnamefont {Liu}},\ }\href
  {\doibase 10.1007/JHEP03(2018)070} {\bibfield  {journal} {\bibinfo  {journal}
  {JHEP}\ }\textbf {\bibinfo {volume} {03}},\ \bibinfo {pages} {070} (\bibinfo
  {year} {2018}{\natexlab{b}})},\ \Eprint {http://arxiv.org/abs/1710.10458}
  {arXiv:1710.10458 [hep-th]} \BibitemShut {NoStop}%
\bibitem [{\citenamefont {Dymarsky}\ and\ \citenamefont
  {Pavlenko}(2019)}]{Dymarsky:2019etq}%
  \BibitemOpen
  \bibfield  {author} {\bibinfo {author} {\bibfnamefont {A.}~\bibnamefont
  {Dymarsky}}\ and\ \bibinfo {author} {\bibfnamefont {K.}~\bibnamefont
  {Pavlenko}},\ }\href {\doibase 10.1103/PhysRevLett.123.111602} {\bibfield
  {journal} {\bibinfo  {journal} {Phys. Rev. Lett.}\ }\textbf {\bibinfo
  {volume} {123}},\ \bibinfo {pages} {111602} (\bibinfo {year} {2019})},\
  \Eprint {http://arxiv.org/abs/1903.03559} {arXiv:1903.03559 [hep-th]}
  \BibitemShut {NoStop}%
\bibitem [{Note1()}]{Note1}%
  \BibitemOpen
  \bibinfo {note} {As explained in Section 2, ${\Delta x^-}$ is the separation
  between the insertions of the light operators in the rescaled lightlike
  direction.}\BibitemShut {Stop}%
\bibitem [{\citenamefont {Fitzpatrick}\ \emph {et~al.}(2013)\citenamefont
  {Fitzpatrick}, \citenamefont {Kaplan}, \citenamefont {Poland},\ and\
  \citenamefont {Simmons-Duffin}}]{Fitzpatrick:2012yx}%
  \BibitemOpen
  \bibfield  {author} {\bibinfo {author} {\bibfnamefont {A.}~\bibnamefont
  {Fitzpatrick}}, \bibinfo {author} {\bibfnamefont {J.}~\bibnamefont {Kaplan}},
  \bibinfo {author} {\bibfnamefont {D.}~\bibnamefont {Poland}}, \ and\ \bibinfo
  {author} {\bibfnamefont {D.}~\bibnamefont {Simmons-Duffin}},\ }\href
  {\doibase 10.1007/JHEP12(2013)004} {\bibfield  {journal} {\bibinfo  {journal}
  {JHEP}\ }\textbf {\bibinfo {volume} {12}},\ \bibinfo {pages} {004} (\bibinfo
  {year} {2013})},\ \Eprint {http://arxiv.org/abs/1212.3616} {arXiv:1212.3616
  [hep-th]} \BibitemShut {NoStop}%
\bibitem [{\citenamefont {Komargodski}\ and\ \citenamefont
  {Zhiboedov}(2013)}]{Komargodski:2012ek}%
  \BibitemOpen
  \bibfield  {author} {\bibinfo {author} {\bibfnamefont {Z.}~\bibnamefont
  {Komargodski}}\ and\ \bibinfo {author} {\bibfnamefont {A.}~\bibnamefont
  {Zhiboedov}},\ }\href {\doibase 10.1007/JHEP11(2013)140} {\bibfield
  {journal} {\bibinfo  {journal} {JHEP}\ }\textbf {\bibinfo {volume} {11}},\
  \bibinfo {pages} {140} (\bibinfo {year} {2013})},\ \Eprint
  {http://arxiv.org/abs/1212.4103} {arXiv:1212.4103 [hep-th]} \BibitemShut
  {NoStop}%
\bibitem [{\citenamefont {Fitzpatrick}\ \emph {et~al.}(2014)\citenamefont
  {Fitzpatrick}, \citenamefont {Kaplan},\ and\ \citenamefont
  {Walters}}]{Fitzpatrick:2014vua}%
  \BibitemOpen
  \bibfield  {author} {\bibinfo {author} {\bibfnamefont {A.~L.}\ \bibnamefont
  {Fitzpatrick}}, \bibinfo {author} {\bibfnamefont {J.}~\bibnamefont {Kaplan}},
  \ and\ \bibinfo {author} {\bibfnamefont {M.~T.}\ \bibnamefont {Walters}},\
  }\href {\doibase 10.1007/JHEP08(2014)145} {\bibfield  {journal} {\bibinfo
  {journal} {JHEP}\ }\textbf {\bibinfo {volume} {08}},\ \bibinfo {pages} {145}
  (\bibinfo {year} {2014})},\ \Eprint {http://arxiv.org/abs/1403.6829}
  {arXiv:1403.6829 [hep-th]} \BibitemShut {NoStop}%
\bibitem [{\citenamefont {Fitzpatrick}\ \emph {et~al.}(2015)\citenamefont
  {Fitzpatrick}, \citenamefont {Kaplan},\ and\ \citenamefont
  {Walters}}]{Fitzpatrick:2015zha}%
  \BibitemOpen
  \bibfield  {author} {\bibinfo {author} {\bibfnamefont {A.~L.}\ \bibnamefont
  {Fitzpatrick}}, \bibinfo {author} {\bibfnamefont {J.}~\bibnamefont {Kaplan}},
  \ and\ \bibinfo {author} {\bibfnamefont {M.~T.}\ \bibnamefont {Walters}},\
  }\href {\doibase 10.1007/JHEP11(2015)200} {\bibfield  {journal} {\bibinfo
  {journal} {JHEP}\ }\textbf {\bibinfo {volume} {11}},\ \bibinfo {pages} {200}
  (\bibinfo {year} {2015})},\ \Eprint {http://arxiv.org/abs/1501.05315}
  {arXiv:1501.05315 [hep-th]} \BibitemShut {NoStop}%
\bibitem [{\citenamefont {Hijano}\ \emph
  {et~al.}(2015{\natexlab{a}})\citenamefont {Hijano}, \citenamefont {Kraus},\
  and\ \citenamefont {Snively}}]{Hijano:2015rla}%
  \BibitemOpen
  \bibfield  {author} {\bibinfo {author} {\bibfnamefont {E.}~\bibnamefont
  {Hijano}}, \bibinfo {author} {\bibfnamefont {P.}~\bibnamefont {Kraus}}, \
  and\ \bibinfo {author} {\bibfnamefont {R.}~\bibnamefont {Snively}},\ }\href
  {\doibase 10.1007/JHEP07(2015)131} {\bibfield  {journal} {\bibinfo  {journal}
  {JHEP}\ }\textbf {\bibinfo {volume} {07}},\ \bibinfo {pages} {131} (\bibinfo
  {year} {2015}{\natexlab{a}})},\ \Eprint {http://arxiv.org/abs/1501.02260}
  {arXiv:1501.02260 [hep-th]} \BibitemShut {NoStop}%
\bibitem [{\citenamefont {Hijano}\ \emph
  {et~al.}(2015{\natexlab{b}})\citenamefont {Hijano}, \citenamefont {Kraus},
  \citenamefont {Perlmutter},\ and\ \citenamefont {Snively}}]{Hijano:2015qja}%
  \BibitemOpen
  \bibfield  {author} {\bibinfo {author} {\bibfnamefont {E.}~\bibnamefont
  {Hijano}}, \bibinfo {author} {\bibfnamefont {P.}~\bibnamefont {Kraus}},
  \bibinfo {author} {\bibfnamefont {E.}~\bibnamefont {Perlmutter}}, \ and\
  \bibinfo {author} {\bibfnamefont {R.}~\bibnamefont {Snively}},\ }\href
  {\doibase 10.1007/JHEP12(2015)077} {\bibfield  {journal} {\bibinfo  {journal}
  {JHEP}\ }\textbf {\bibinfo {volume} {12}},\ \bibinfo {pages} {077} (\bibinfo
  {year} {2015}{\natexlab{b}})},\ \Eprint {http://arxiv.org/abs/1508.04987}
  {arXiv:1508.04987 [hep-th]} \BibitemShut {NoStop}%
\bibitem [{\citenamefont {Fitzpatrick}\ \emph {et~al.}(2016)\citenamefont
  {Fitzpatrick}, \citenamefont {Kaplan}, \citenamefont {Walters},\ and\
  \citenamefont {Wang}}]{Fitzpatrick:2015foa}%
  \BibitemOpen
  \bibfield  {author} {\bibinfo {author} {\bibfnamefont {A.~L.}\ \bibnamefont
  {Fitzpatrick}}, \bibinfo {author} {\bibfnamefont {J.}~\bibnamefont {Kaplan}},
  \bibinfo {author} {\bibfnamefont {M.~T.}\ \bibnamefont {Walters}}, \ and\
  \bibinfo {author} {\bibfnamefont {J.}~\bibnamefont {Wang}},\ }\href {\doibase
  10.1007/JHEP05(2016)069} {\bibfield  {journal} {\bibinfo  {journal} {JHEP}\
  }\textbf {\bibinfo {volume} {05}},\ \bibinfo {pages} {069} (\bibinfo {year}
  {2016})},\ \Eprint {http://arxiv.org/abs/1510.00014} {arXiv:1510.00014
  [hep-th]} \BibitemShut {NoStop}%
\bibitem [{\citenamefont {Cotler}\ and\ \citenamefont
  {Jensen}(2019)}]{Cotler:2018zff}%
  \BibitemOpen
  \bibfield  {author} {\bibinfo {author} {\bibfnamefont {J.}~\bibnamefont
  {Cotler}}\ and\ \bibinfo {author} {\bibfnamefont {K.}~\bibnamefont
  {Jensen}},\ }\href {\doibase 10.1007/JHEP02(2019)079} {\bibfield  {journal}
  {\bibinfo  {journal} {JHEP}\ }\textbf {\bibinfo {volume} {02}},\ \bibinfo
  {pages} {079} (\bibinfo {year} {2019})},\ \Eprint
  {http://arxiv.org/abs/1808.03263} {arXiv:1808.03263 [hep-th]} \BibitemShut
  {NoStop}%
\bibitem [{\citenamefont {Collier}\ \emph {et~al.}(2018)\citenamefont
  {Collier}, \citenamefont {Gobeil}, \citenamefont {Maxfield},\ and\
  \citenamefont {Perlmutter}}]{Collier:2018exn}%
  \BibitemOpen
  \bibfield  {author} {\bibinfo {author} {\bibfnamefont {S.}~\bibnamefont
  {Collier}}, \bibinfo {author} {\bibfnamefont {Y.}~\bibnamefont {Gobeil}},
  \bibinfo {author} {\bibfnamefont {H.}~\bibnamefont {Maxfield}}, \ and\
  \bibinfo {author} {\bibfnamefont {E.}~\bibnamefont {Perlmutter}},\
  }\href@noop {} {\  (\bibinfo {year} {2018})},\ \Eprint
  {http://arxiv.org/abs/1811.05710} {arXiv:1811.05710 [hep-th]} \BibitemShut
  {NoStop}%
\bibitem [{\citenamefont {Maxfield}(2017)}]{Maxfield:2017rkn}%
  \BibitemOpen
  \bibfield  {author} {\bibinfo {author} {\bibfnamefont {H.}~\bibnamefont
  {Maxfield}},\ }\href@noop {} {\  (\bibinfo {year} {2017})},\ \Eprint
  {http://arxiv.org/abs/1712.00885} {arXiv:1712.00885 [hep-th]} \BibitemShut
  {NoStop}%
\bibitem [{Note2()}]{Note2}%
  \BibitemOpen
  \bibinfo {note} {In the following we mostly consider $d>2$, but the $d=2$
  limit can also be recovered.}\BibitemShut {Stop}%
\bibitem [{\citenamefont {Huang}(2019)}]{Huang:2019fog}%
  \BibitemOpen
  \bibfield  {author} {\bibinfo {author} {\bibfnamefont {K.-W.}\ \bibnamefont
  {Huang}},\ }\href {\doibase 10.1103/PhysRevD.100.061701} {\bibfield
  {journal} {\bibinfo  {journal} {Phys. Rev. D}\ }\textbf {\bibinfo {volume}
  {100}},\ \bibinfo {pages} {061701} (\bibinfo {year} {2019})},\ \Eprint
  {http://arxiv.org/abs/1907.00599} {arXiv:1907.00599 [hep-th]} \BibitemShut
  {NoStop}%
\bibitem [{\citenamefont {Huang}(2020)}]{Huang:2020ycs}%
  \BibitemOpen
  \bibfield  {author} {\bibinfo {author} {\bibfnamefont {K.-W.}\ \bibnamefont
  {Huang}},\ }\href@noop {} {\  (\bibinfo {year} {2020})},\ \Eprint
  {http://arxiv.org/abs/2002.00110} {arXiv:2002.00110 [hep-th]} \BibitemShut
  {NoStop}%
\bibitem [{Note3()}]{Note3}%
  \BibitemOpen
  \bibinfo {note} {Note that multi stress tensor operators with any derivatives
  in them do not contribute in the large $R$ limit.}\BibitemShut {Stop}%
\bibitem [{\citenamefont {Bianchi}\ \emph {et~al.}(2020)\citenamefont
  {Bianchi}, \citenamefont {Grillo},\ and\ \citenamefont
  {Morales}}]{Bianchi:2020des}%
  \BibitemOpen
  \bibfield  {author} {\bibinfo {author} {\bibfnamefont {M.}~\bibnamefont
  {Bianchi}}, \bibinfo {author} {\bibfnamefont {A.}~\bibnamefont {Grillo}}, \
  and\ \bibinfo {author} {\bibfnamefont {J.~F.}\ \bibnamefont {Morales}},\
  }\href@noop {} {\  (\bibinfo {year} {2020})},\ \Eprint
  {http://arxiv.org/abs/2002.05574} {arXiv:2002.05574 [hep-th]} \BibitemShut
  {NoStop}%
\bibitem [{\citenamefont {Brigante}\ \emph
  {et~al.}(2008{\natexlab{a}})\citenamefont {Brigante}, \citenamefont {Liu},
  \citenamefont {Myers}, \citenamefont {Shenker},\ and\ \citenamefont
  {Yaida}}]{Brigante:2007nu}%
  \BibitemOpen
  \bibfield  {author} {\bibinfo {author} {\bibfnamefont {M.}~\bibnamefont
  {Brigante}}, \bibinfo {author} {\bibfnamefont {H.}~\bibnamefont {Liu}},
  \bibinfo {author} {\bibfnamefont {R.~C.}\ \bibnamefont {Myers}}, \bibinfo
  {author} {\bibfnamefont {S.}~\bibnamefont {Shenker}}, \ and\ \bibinfo
  {author} {\bibfnamefont {S.}~\bibnamefont {Yaida}},\ }\href {\doibase
  10.1103/PhysRevD.77.126006} {\bibfield  {journal} {\bibinfo  {journal} {Phys.
  Rev. D}\ }\textbf {\bibinfo {volume} {77}},\ \bibinfo {pages} {126006}
  (\bibinfo {year} {2008}{\natexlab{a}})},\ \Eprint
  {http://arxiv.org/abs/0712.0805} {arXiv:0712.0805 [hep-th]} \BibitemShut
  {NoStop}%
\bibitem [{\citenamefont {Brigante}\ \emph
  {et~al.}(2008{\natexlab{b}})\citenamefont {Brigante}, \citenamefont {Liu},
  \citenamefont {Myers}, \citenamefont {Shenker},\ and\ \citenamefont
  {Yaida}}]{Brigante:2008gz}%
  \BibitemOpen
  \bibfield  {author} {\bibinfo {author} {\bibfnamefont {M.}~\bibnamefont
  {Brigante}}, \bibinfo {author} {\bibfnamefont {H.}~\bibnamefont {Liu}},
  \bibinfo {author} {\bibfnamefont {R.~C.}\ \bibnamefont {Myers}}, \bibinfo
  {author} {\bibfnamefont {S.}~\bibnamefont {Shenker}}, \ and\ \bibinfo
  {author} {\bibfnamefont {S.}~\bibnamefont {Yaida}},\ }\href {\doibase
  10.1103/PhysRevLett.100.191601} {\bibfield  {journal} {\bibinfo  {journal}
  {Phys. Rev. Lett.}\ }\textbf {\bibinfo {volume} {100}},\ \bibinfo {pages}
  {191601} (\bibinfo {year} {2008}{\natexlab{b}})},\ \Eprint
  {http://arxiv.org/abs/0802.3318} {arXiv:0802.3318 [hep-th]} \BibitemShut
  {NoStop}%
\bibitem [{\citenamefont {Hofman}\ and\ \citenamefont
  {Maldacena}(2008)}]{Hofman:2008ar}%
  \BibitemOpen
  \bibfield  {author} {\bibinfo {author} {\bibfnamefont {D.~M.}\ \bibnamefont
  {Hofman}}\ and\ \bibinfo {author} {\bibfnamefont {J.}~\bibnamefont
  {Maldacena}},\ }\href {\doibase 10.1088/1126-6708/2008/05/012} {\bibfield
  {journal} {\bibinfo  {journal} {JHEP}\ }\textbf {\bibinfo {volume} {05}},\
  \bibinfo {pages} {012} (\bibinfo {year} {2008})},\ \Eprint
  {http://arxiv.org/abs/0803.1467} {arXiv:0803.1467 [hep-th]} \BibitemShut
  {NoStop}%
\end{thebibliography}%

\pagebreak

\appendix

\titlepage

\setcounter{page}{1}

\begin{center}
{\LARGE Supplemental Material}
\end{center}

\section{A. Details of the $d=2$ case}
In two spacetime dimensions some simplifications occur.
Eq. (\ref{alphasol}) can be written as
\be
\label{alphasoltwo}
  \sqrt{\mu} \dxp = 4 \tanh \sqrt{ \tilde x_1}, \qquad \tilde x_1 \equiv \frac{x_1}{x_0} = \frac{\alpha}{x_0^2} ,
\ee
where $x_{1,2}$ are the roots of the polynomial $P_2(x) = x^2 - 4 x +\alpha$ in the decreasing order.
Note that $x_0 = 4/(1+\tilde x_1)$.
Combining everything yields
\be
\label{elltwo}
  \ell_f|_{d=2} = \log\frac{\sqrt{\tilde x_1}}{1-\tilde x_1} ,
 \ee
which, upon the substitution of (\ref{alphasoltwo}) results in eq. (\ref{dtwo}).

\section{B. Details of the $d=4$ case}

The integrals can be  rewritten in terms of  special functions.
 For example,
 \be
 \label{iplus}
    I_+^{(d=4)} = \frac{4}{x_0} \,  F_1(1,\frac{1}{2},\frac{1}{2},\frac{3}{2},\tilde x_1,\tilde x_2) ,
 \ee
 where $\tilde x_{1,2} = x_{1,2}/x_0$,  $x_i, \; i=0,1,2$ are the ($\alpha$-dependent) roots of the polynomial
 $P_4(x) = x^3- 4 x^2 + \alpha$ in the decreasing order and $F_1$ is the Appel function.
 Likewise,
 \be
 \label{iminus}
    I_-^{(d=4)}  = \frac{4}{x_0} \,  F_1(1,\frac{1}{2},\frac{1}{2},\frac{3}{2},\tilde x_1,\tilde x_2 ) \\
          - \frac{16 \alpha}{15 x_0^3} \,  F_1(3,\frac{1}{2},\frac{1}{2},\frac{7}{2},\tilde x_1,\tilde x_2) .
 \ee
It will be convenient to parameterize the solutions in terms of a new variable, $\bar \alpha =\alpha/x_0^2$.
One can  derive the following expressions:
\be
\label{xonetwo}
  \tilde x_{1} (\bara) = \frac{ \bara + \sqrt{  (16 - 3 \bara)  \bara}  }{  8- 2 \bara }, \qquad   \tilde x_{2} (\bara) = \frac{ \bara - \sqrt{  (16 - 3 \bara )  \bara}  }{  8- 2 \bara }, \qquad x_0 (\bara)= 4- \bara .
\ee
Hence, eq. (\ref{alphasol}) becomes
\be
\label{solbara}
 \dxm (\dxp)^3  = -\frac{ 16 \bara}{\left( 1 - \frac{\bara}{4}\right)^2} F_1^3(1,\frac{1}{2},\frac{1}{2},\frac{3}{2},\tilde x_1,\tilde x_2) 
                           \left[   F_1(1,\frac{1}{2},\frac{1}{2},\frac{3}{2},\tilde x_1,\tilde x_2)    - \frac{4 \bara}{15} \,  F_1(3,\frac{1}{2},\frac{1}{2},\frac{7}{2},\tilde x_1,\tilde x_2)   \right] ,
\ee
where $x_{1,2} = x_{1,2}(\bara)$.
We generally need to invert this, which is a simple exercise perturbatively,
\be
\label{invertbara}
     \bara= -\frac{\Delta_x}{16} \left( 1+\frac{5 \Delta_x}{96} + \frac{6299 \Delta_x^2}{2150400}   + \frac{76228319 \Delta_x^3}{442810368000} +\ldots \right), \qquad \Delta_x = \dxm (\dxp)^3 .
\ee
To deal with the $\log$ divergence, one can differentiate $I_\ell$ in (\ref{ell_planara}) with respect to $\bara$.
This  makes  the integral convergent and kills the $\log \Lambda_u$ term, but one would need to integrate back with respect to $\bara$
to recover the result:
\be
\label{sol}
  \ell_f|_{d=4} = \log(\dxp \dxm) -\log[ I_+ ^{(d=4)} I_-^{(d=4)} ]+ I_\ell^{(d=4)} (\bara)  ,
\ee
where $I_+^{(d=4)} $ and and $I_-^{(d=4)} $ are given by (\ref{iplus}), (\ref{iminus}) and $\tilde x_{1,2}$ and $x_0$ are related to 
$\bara $ by (\ref{xonetwo}), while $I^{(d=4)}(\bara) $ is given by 
\be
\label{defi}
\begin{split}
 I_\ell^{(d=4)}(\bara)  &= 2 \int_0^\bara \frac{   dt }{(t-4)^2 \sqrt{(16-3 t)t} } 
  \Big[     (8 - t+\sqrt{(16-3 t)t})  F_1[1,\frac{3}{2},\frac{1}{2},\frac{3}{2},\tilde x_1(t),\tilde x_2(t)] \\
 &\qquad +   (t-8+\sqrt{(16-3 t)t})  F_1[1,\frac{1}{2},\frac{3}{2},\frac{3}{2},\tilde x_1(t),\tilde x_2(t)]     \Big] -\log (4-\bara) .
\end{split}
\ee
Eq. (\ref{defi}) can  be expanded perturbatively in $\bara$,
\be
\label{solpert}
    I_\ell^{(d=4)}(\bara)  =\frac{2\bara}{3}+\frac{5\bara^2}{21}+\frac{647\bara^3}{5544}+\frac{391 \bara^4}{6006} +\ldots ,
\ee
where we omit a constant term.
Substituting eq. (\ref{solpert}) into eq. (\ref{sol}) yields eq. (\ref{ellcutoff_f}).

\end{document}